\documentclass[useAMS,usenatbib]{mn2e}
\usepackage[final]{epsfig}
\usepackage{latexsym}
\usepackage{psfrag}

\title[Uniformly rotating axisymmetric fluid configurations]
      {Uniformly rotating axisymmetric fluid configurations
       bifurcating from highly flattened Maclaurin spheroids}
\author[Marcus Ansorg, Andreas Kleinw\"{a}chter and Reinhard
        Meinel]
{ 
  Marcus Ansorg\thanks{E-mail:ansorg@tpi.uni-jena.de}, 
  Andreas Kleinw\"{a}chter
  and Reinhard Meinel\\
  Theoretisch-Physikalisches Institut, University of Jena \\
  Max-Wien-Platz 1 \\ 07743 Jena, Germany
}

\begin{document}
%\date{Accepted 2002 August 15. Received 2002 July 14}

\maketitle

\begin{abstract}
We give a thorough investigation of sequences 
of uniformly rotating, homogeneous axisymmetric Newtonian equilibrium 
configurations that bifurcate from highly flattened Maclaurin spheroids. 
Each one of these sequences possesses a mass-shedding limit. Starting 
at this point, the sequences proceed towards the Maclaurin 
sequence and beyond. The first sequence leads to the well known Dyson rings,
whereas the end points of the higher sequences are characterized by the 
formation of a two-body system, either a core-ring system (for the second, 
the fourth etc.~sequence) or a two-ring system (for the third, the fifth 
etc.~sequence). Although the general qualitative picture drawn by Eriguchi 
and Hachisu in the eighties has been confirmed, slight differences turned out
in the
interpretation of the origin of the first two-ring sequence and in the general 
appearance of fluid bodies belonging to higher sequences.
\end{abstract}


%_________________________
\section{Introduction}
If one moves along the Maclaurin sequence of uniformly rotating, axisymmetric,
homogeneous fluid ellipsoids with fixed mass and fixed density, 
starting at the
non-rotating configuration and proceeding towards increasing angular 
momentum, one first encounters the Jacobi branch point where 
the ellipsoids become secularly unstable with regard to the first 
non-axisymmetric 
perturbation (see e.g.~Chandrasekhar 1969).
The corresponding Jacobi sequence branching off at this point 
leads to further bifurcations, in particular to Poincar\'e's 
`pear-shaped' sequence (see Eriguchi \& Hachisu 1982b).
Farther along the Maclaurin path, one next comes to the bifurcation points 
of the non-axisymmetric `triangle', `square' and 
`ammonite' sequences (Chandrasekhar 1969, Eriguchi \& Hachisu 1982a), 
before one arrives at the first axisymmetric sequence
bifurcating at an eccentricity $\varepsilon_1=0.98523$ (Chandrasekhar 1967,
Bardeen 1971). As conjectured by 
Bardeen (1971) and confirmed by Eriguchi \& Sugimoto (1981), 
the bodies of this sequence pinch together  
gradually at the centre and eventually form the anchor-ring configurations 
studied by Dyson (1892, 1893) and Wong (1974); see also Poincar\'e (1885), 
Kowalewsky (1895), Lichtenstein (1933) and a somewhat related paper by Kley
(1996).
 
Another not yet confirmed conjecture by Bardeen (1971)
concerns a sequence of  axisymmetric `central-bulge' configurations likewise 
bifurcating at $\varepsilon_1$. In this paper we indeed found this sequence,
which might be considered to be a continuation of the Dyson-ring sequence 
beyond the Maclaurin ellipsoids and ends in a mass-shedding limit. 
However, the surface shapes of the 
corresponding fluid bodies do not generate a `central-bulge' region; 
their appearance is more `lens shaped'.

Apart from the study of the Dyson-ring sequence we shall give a detailed
analysis of the next axisymmetric sequences bifurcating from the 
Maclaurin ellipsoids. The first core-ring sequence branches off at 
$\varepsilon_2=0.99375$. Starting at a mass-shedding 
limit, this sequence proceeds towards the Maclaurin sequence and beyond,
finally leading to the formation of a core-ring system.

In contrast to the results by Eriguchi \& Hachisu (1982a) who stated that
the first two-ring sequence
branches off at $\varepsilon_2$, we found that the bifurcation occurs at 
$\varepsilon_3=0.99657$. Again, a mass-shedding limit marks one end point 
of this sequence, leading from here towards and beyond the 
Maclaurin sequence and ending in the formation of a two-ring system.

The same qualitative picture repeats as one moves to the higher axisymmetric
bifurcation points at $\varepsilon_4=0.99784,
\varepsilon_5=0.99851$ etc., of which there are infinitely many, 
accumulating at $\varepsilon=1$ (Bardeen 1971). 
Starting at a mass-shedding limit, the sequences proceed towards the 
Maclaurin sequence and beyond, leading eventually to the 
formation of a core-ring system for $\varepsilon_{2l}$ 
and a two-ring system for $\varepsilon_{2l+1}$, see Figs 7-11. 
The body's surface is characterized by a particular number
of grooves, notably $l$, for configurations close to the
two-body systems that are the end-stages of the 
sequences bifurcating at $\varepsilon_{2l}$ and $\varepsilon_{2l+1}$. The
outermost groove pinches together first and an outer ring without grooves 
separates. 
In some continuation process beyond this formation of a two-body system, the
other grooves might also pinch off, leading eventually to a multi-body system, 
at most consisting of a core and $l$ rings or $(l+1)$ rings 
respectively\footnote{This result 
is in contrast to the statement by Eriguchi \& Hachisu (1982a), 
who expected the 
formation of $k$ rings for the $\varepsilon_k$-sequence.}.
However, since a continuation process of this kind is not unique, we consider
the sequences in question precisely up to the formation of 
the two-body system.

These investigations have been carried out up to the
$\varepsilon_{10}$-sequence. The corresponding final configurations, 
when the two-body system is formed, can be seen in Figs 7-11. 

Representative plots of physical quantities as well as of meridional 
cross-sections have been provided
up to the $\varepsilon_{5}$-sequence, see Figs 1-10. 
Note that the Maclaurin sequence becomes dynamically unstable with respect
to axisymmetric perturbations for $\varepsilon>0.99856$ (Bardeen 1971, 
Eriguchi \& Hachisu 1985).
Therefore, if we restrict our considerations to
axial symmetry, the sequences bifurcating at
$\varepsilon_1,\varepsilon_2,\ldots,\varepsilon_5$ are more relevant than
those branching off at $\varepsilon_6,\varepsilon_7,\ldots\,
(\varepsilon_k\geq\varepsilon_6=0.99891$ for $k>5)$.

Table 1 lists physical quantities for the
Maclaurin ellipsoids at the bifurcation points 
$\varepsilon_k$ for $k=1,2,\ldots,10$. Tables 2-4 contain numerical 
data with an accuracy of five digits for configurations of the 
$\varepsilon_{1}$, $\varepsilon_{2}$ and $\varepsilon_{3}$-sequences. 
Additionally, Table 5 lists the ratios of the masses of the inner to the 
outer body at the two-body formation point for the
$\varepsilon_2\ldots\varepsilon_5$-sequences.

\begin{table}{{\bf Table 1:} Physical quantities for the
Maclaurin ellipsoids at the bifurcation points 
$\varepsilon_k$. For the definition of $\omega^2$, $j^2$, $T$ and $W$ see
Section 4.}
\begin{center}
\begin{tabular}[b]{ccccccl}
\hline\hline
$k$ & $\varepsilon_k$ & Axis ratio & $\omega^2$ & $j^2$    & $T/|W|$  \\
\hline
 1 & 0.98523 & 0.17126 & 0.087262 & 0.021741 & 0.35890 \\
 2 & 0.99375 & 0.11160 & 0.066105 & 0.029152 & 0.40345 \\
 3 & 0.99657 & 0.082750 & 0.052711 & 0.034638 & 0.42664 \\
 4 & 0.99784 & 0.065744 & 0.043714 & 0.039037 & 0.44084 \\
 5 & 0.99851 & 0.054534 & 0.037301 & 0.042740 & 0.45044 \\
 6 & 0.99891 & 0.046589 & 0.032513 & 0.045957 & 0.45736 \\
 7 & 0.99917 & 0.040664 & 0.028806 & 0.048814 & 0.46259 \\
 8 & 0.99935 & 0.036075 & 0.025854 & 0.051395 & 0.46667  \\
 9 & 0.99947 & 0.032417 & 0.023449 & 0.053755 & 0.46995 \\
10 & 0.99957 & 0.029432 & 0.021451 & 0.055935 & 0.47265
 \end{tabular}
\end{center}
\end{table}

\begin{figure*}[h]
\unitlength1cm
\hspace*{-2cm}
\epsfig{file=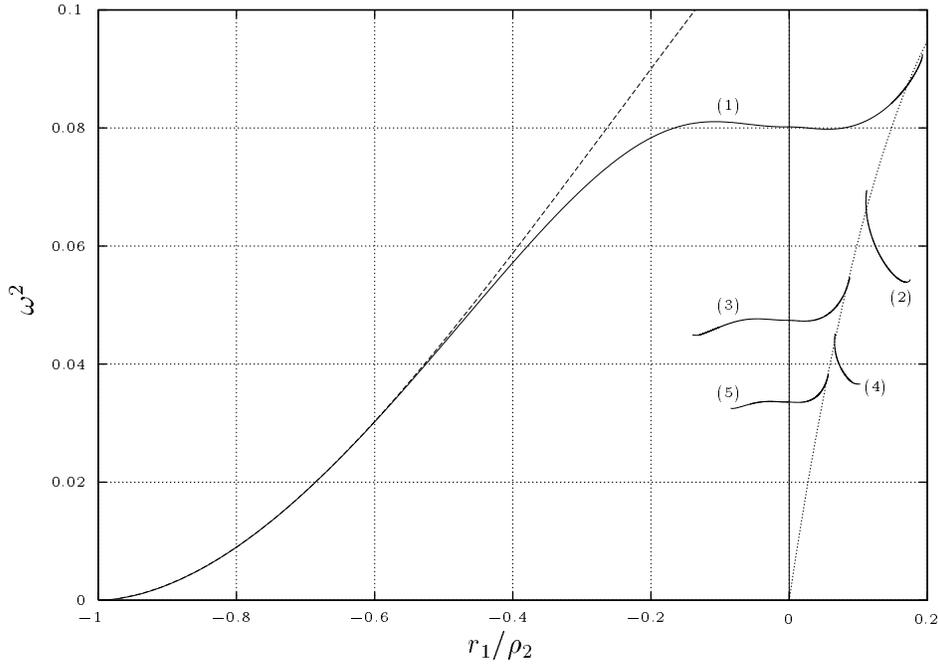,scale=1}
\vspace*{-17cm}
\caption{For the first five axisymmetric sequences $(1), \ldots,(5)\,,$
the squared angular velocity $\omega^2=\Omega^2/(4\pi G\mu)$
is plotted against the radius ratio $r_1/\rho_2$, where $r_1=\zeta_1$ 
($\zeta_1:$ polar radius) for spheroidal figures and $r_1=-\rho_1$
($\rho_1:$ inner equatorial radius) for toroidal shapes; $\rho_2$ 
is always the radius of the outer equatorial rim. 
The dotted curve represents the Maclaurin sequence and 
the dashed one corresponds to the Dyson approximation.}
\end{figure*}

\begin{figure*}[h]
\unitlength1cm
\hspace*{-2cm}
\epsfig{file=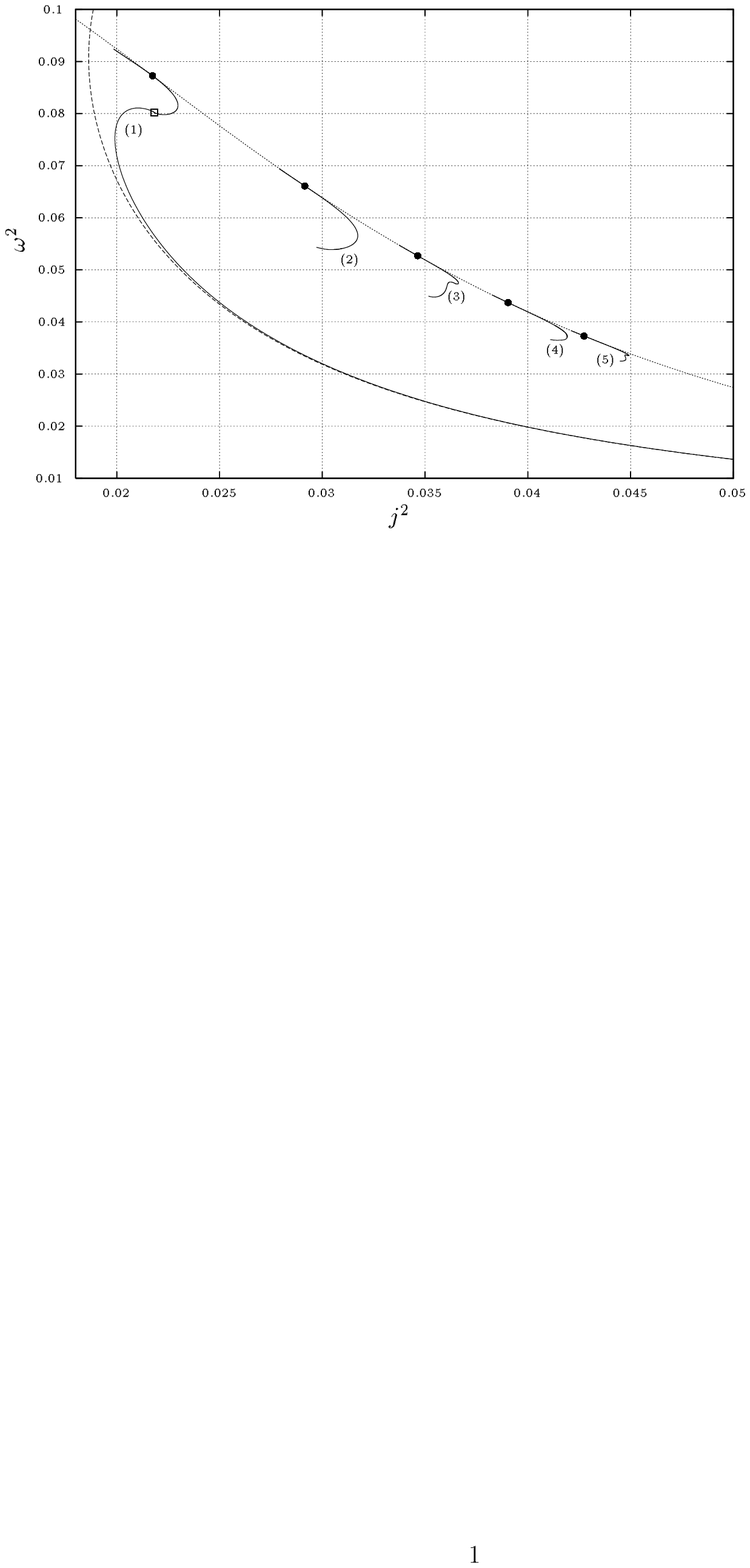,scale=1}
\vspace*{-17cm}
\caption{For the first five axisymmetric sequences, 
the squared angular velocity $\omega^2=\Omega^2/(4\pi G\mu)$
is plotted against the dimensionless squared angular momentum $j^2$, given in
formula (\ref{j2}). Dotted and dashed curves again refer to the Maclaurin
sequence and the Dyson approximation respectively. The $\bullet$'s mark the
bifurcation points on the Maclaurin sequence and $\Box$ the transition
configuration of spheroidal to toroidal bodies on the Dyson-ring sequence.}
\end{figure*}

%_________________________
\section{Bifurcation points of axisymmetric sequences}
In this Section we review the results by Bardeen (1971) who obtained all
bifurcation points on the Maclaurin sequence which correspond to axisymmetric 
perturbations. 
For our purposes, we will use the Maclaurin ellipsoids at these points 
as initial solutions
for finding the respective sequence branching off. 

The solution $\Phi$ of the axisymmetric Poisson equation 
   \begin{equation}
   \label{Poisson}
   \triangle \Phi=4\pi\;G\mu
   \end{equation}
where $G$ is Newton's constant of gravity, $\mu$ the mass density, and
$\triangle$ the Laplace operator, 
given in cylindrical coordinates $(\rho,\zeta)$ by
   \[\triangle=\frac{\partial^2}{\partial\rho^2}+\frac{1}{\rho}
   \frac{\partial}{\partial\rho}+\frac{\partial^2}{\partial\zeta^2}\,,\]
can be written in the form (Morse \& Feshbach 1953)
\begin{eqnarray*}
\lefteqn{
\Phi(\xi,\eta)=-2\pi a_0^2\,G\sum_{l=0}^\infty
(2l+1)P_l(\eta)\times}\\&&\times
\left[p_l(\xi)\int\limits_\xi^\infty\int\limits_{-1}^1
(\xi'^{\,2}+\eta'^{\,2})P_l(\eta')q_l(\xi')\mu(\xi',\eta')d\eta'
d\xi'\right.\\&&
+\left.q_l(\xi)\int\limits_0^\xi\int\limits_{-1}^1
(\xi'^{\,2}+\eta'^{\,2})P_l(\eta')p_l(\xi')\mu(\xi',\eta')d\eta' d\xi'\right].
\end{eqnarray*}
Here, the oblate spheroidal coordinates $\xi$ and $\eta$ are connected with
$\rho$ and $\zeta$ by
\[\rho=a_0\sqrt{(1+\xi^2)(1-\eta^2)}\;,\quad\zeta=a_0\xi\eta\,.\] 
The functions $P_l$ are Legendre polynomials,
and the functions $p_l$ and $q_l$ are defined by
\[p_l(\xi)=(-\mbox{i})^lP_l(\mbox{i}\xi)\;,\quad 
  q_l(\xi)=p_l(\xi)\int_\xi^\infty\frac{d\xi'}{(1+\xi'^{\,2})[p_l(\xi')]^2}\;.
\]
The fluid body revolving with the uniform angular velocity $\Omega$ 
is in hydrostatic equilibrium if the condition
\begin{equation}
\label{OFB}
\Phi(\xi,\eta)-\frac{1}{2}\Omega^2\rho^2=\Phi_0=\mbox{const.}
\end{equation}
is satisfied at every point of the body's surface. 

In order to derive the Maclaurin solutions, one simply starts with an 
ellipsoidal shape of the surface characterized by
$\xi_S(\eta)=\xi_0=\mbox{const.}$ and moreover takes $a_0$ to be the 
corresponding 
focal length. The density $\mu$ vanishes for $\xi>\xi_0$ and is a positive
constant if $\xi<\xi_0$ which in the following we simply call $\mu$. 
Then the resulting surface potential reads as follows
\[\Phi(\xi_0,\eta)=-\frac{4}{3}\pi
a_0^2\,G\,\mu\,\xi_0(1+\xi_0^2)[q_0(\xi_0)+q_2(\xi_0)P_2(\eta)],\]
and the surface condition (\ref{OFB}) indeed is satisfied, provided
that $\Omega$ and $\Phi_0$ are well-defined function of $a_0,\xi_0$ and $\mu$
which can be read off from (\ref{OFB}). 

If we now consider an axisymmetric perturbation of the above Maclaurin
shape
\[\xi_S(\eta)=\xi_0+\kappa\xi_1(\eta)+{\cal O}(\kappa^2)\;,\]
insert the corresponding surface potential into the condition (\ref{OFB}) 
and linearize the resulting expression with respect to the small parameter 
$\kappa$, we uniquely find a neighbouring Maclaurin ellipsoid unless 
\begin{equation}\label{eps_k}
2p_{2k+2}(\xi_0)q_{2k+2}(\xi_0)=
\frac{\xi_0}{p_2(\xi_0)}\left[1-2\xi_0 q_2(\xi_0)\right]
\end{equation}
is valid for some $k\in\{1,2,3,\ldots\}$. The corresponding eccentricity
$\varepsilon_k$ belonging to the root 
$\xi_0^{(k)}$ of the above equation is given by 
\[\varepsilon_{k}=\frac{1}{\sqrt{1+(\xi_0^{(k)})^2}}\;,\quad k=1,2,3,\ldots\]
and marks the bifurcation point on the Maclaurin sequence at which the
$k^{\rm{th}}$
axisymmetric sequence branches off. The numerical values of the first ten
$\varepsilon_k$ are given in Table 1. As illustrated in Section 4, 
the bifurcation points appear as cross-over points of the respective 
sequence with the Maclaurin sequence if one plots the evolution of 
these sequences in an appropriate parameter diagram (such as angular 
velocity against the ratio of polar and equatorial radii). Note that the
corresponding linear surface displacement $\xi_1^{(k)}$ turns out to be
\begin{equation}\label{surf_dis}
\xi_1^{(k)}(\eta)\propto\frac{P_{2k+2}(\eta)}{(\xi_0^{(k)})^2+\eta^2}\;.
\end{equation}

\begin{figure}
\unitlength1cm
\hspace*{5mm}
\epsfig{file=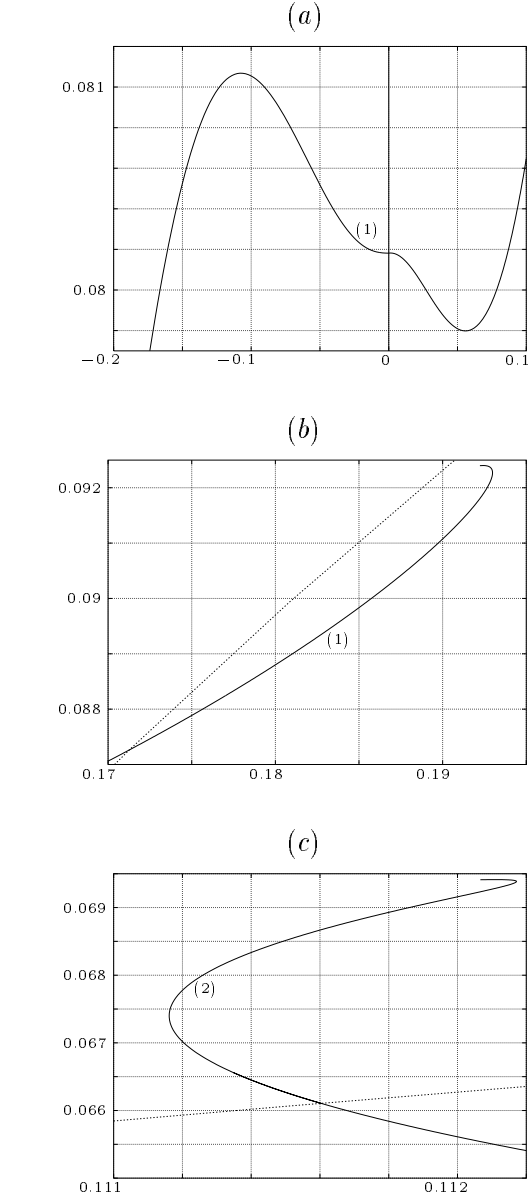,scale=1}
%\vspace*{5mm}
\caption{Magnified details of Fig. 1. The axes of the abscissae and ordinates
have been stretched by a different factor.}
\end{figure}
\begin{figure}
\unitlength1cm
\hspace*{5mm}
\epsfig{file=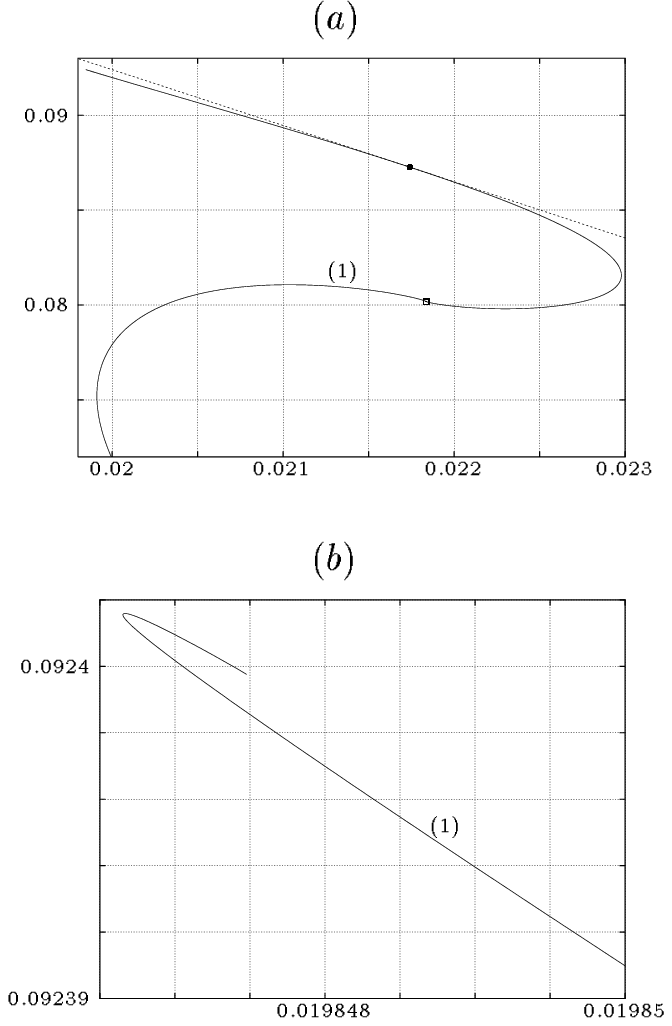,scale=1}
%\vspace*{5mm}
\caption{Magnified details of Fig. 2. The axes of the abscissae and ordinates
have been stretched by a different factor.}
\end{figure}

\begin{figure}
\unitlength1cm
\hspace*{5mm}
\epsfig{file=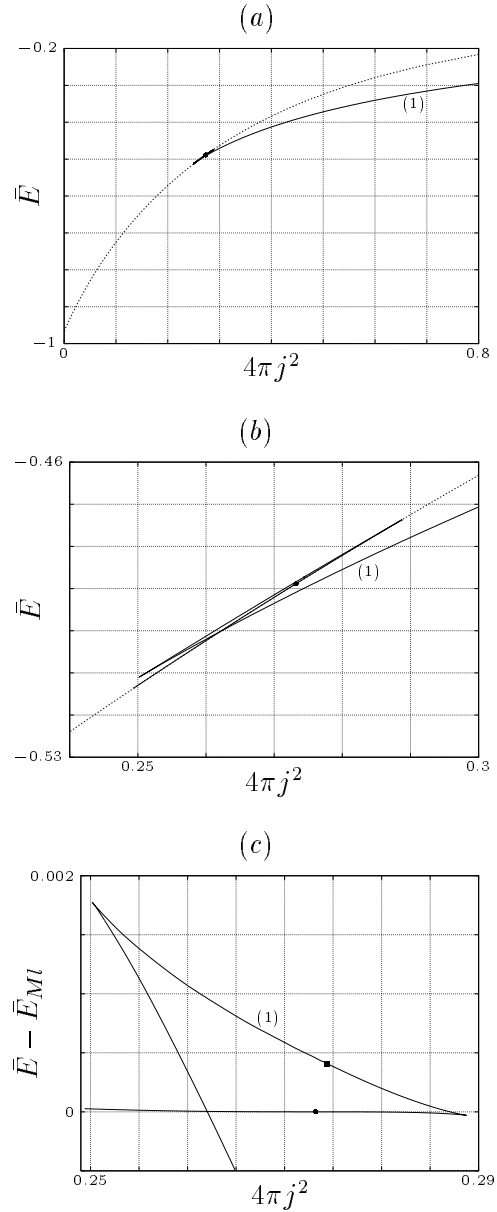,scale=1}
%\vspace*{5mm}
\caption{For the Dyson-ring sequence, 
the dimensionless total energy $\bar E$, given in formula (\ref{barE}), is 
plotted against $4\pi j^2$, see (\ref{j2}). The Maclaurin sequence is
indicated
by the dotted curve; the $\bullet$ marks the corresponding bifurcation point. 
The diagram (b) shows a magnified detail of (a), again with different
stretching
factors for the two axes. In (c), 
the difference $(\bar{E}-\bar{E}_{Ml})$ of the respective energies
of Dyson-ring configurations and Maclaurin solutions belonging to the same 
$j^2$ is displayed. The $\Box$ marks the transition
configuration of spheroidal to toroidal bodies.}
\end{figure}

%_________________________
\section{Numerical methods}
The Poisson-integral
\[\Phi({\bf r})=-G\mu\int\limits_V\frac{d^3{\bf r'}}{|{\bf r}-{\bf r'}|}\]
over the fluid body $V$ can be reduced to a one-dimensional integral 
as follows (Hill \& Wheeler 1953, Wong 1974):
\begin{eqnarray}
\lefteqn{\Phi(\rho,\zeta)}\nonumber\\&&\hspace*{-5mm}\nonumber
=-G\mu\int\limits_{\rho_1}^{\rho_2}\int
\limits_{-\zeta_s(\rho')}^{\zeta_s(\rho')}
\int\limits_0^{2\pi}\frac{\rho'd\phi'd\zeta'd\rho'}
{\sqrt{\rho^2+\rho'^{\,2}+(\zeta-\zeta')^2-2\rho\rho'\cos\phi'}} 
\\&&\hspace*{-5mm}
\label{Pois_Int2}
=-4G\mu\int\limits_{\rho_1}^{\rho_2}\int
\limits_{-\zeta_s(\rho')}^{\zeta_s(\rho')}
\frac{\rho'K(m)d\zeta'd\rho'}
{\sqrt{(\rho+\rho')^2+(\zeta-\zeta')^2}} \\&&\hspace*{-5mm}
\label{Pois_Int3}
=-2G\mu\int\limits_{\rho_1}^{\rho_2}
[A_-(\rho';\rho,\zeta)-A_+(\rho';\rho,\zeta)]\rho'd\rho'
\end{eqnarray}
with
\[A_\pm(\rho';\rho,\zeta)=\]

\smallskip

\[\hspace{1mm}\frac{
\pm[(\rho+\rho')K(m_\pm)-2\rho D(m_\pm)]\zeta'_s(\rho')
+[\zeta\mp\zeta_s(\rho')]K(m_\pm)}
{\sqrt{[\rho+\rho']^2+[\zeta\mp\zeta_s(\rho')]^2}}.\] 

In these expressions the surface is described by a non-negative function 
$\zeta_s$ defined on the interval $[\rho_1,\rho_2]$ and $\zeta_s'$ denotes its
first derivative. Note that $\rho_1=0$ for
a spheroidal equlibrium figure, whereas for a toroidal figure $\rho_1$ denotes
the inner equatorial radius; $\rho_2$ 
is always the radius of the outer equatorial rim. 

The functions $K$ and $D$ 
are defined by the elliptic integrals
\begin{eqnarray*}
  K(m)&=&\int\limits_0^{\pi/2}\frac{d\alpha}{\sqrt{1-m^2\sin^2\alpha}}\;,\\
  D(m)&=&\int\limits_0^{\pi/2}\frac{\sin^2\alpha
  \,d\alpha}{\sqrt{1-m^2\sin^2\alpha}}\;,\end{eqnarray*}
and the modulus is given by
\begin{eqnarray*}
m&=&\sqrt{\frac{4\rho\rho'}{(\rho+\rho')^2+(\zeta-\zeta')^2}}
\qquad\mbox{and}\\
m_\pm&=&\sqrt{\frac{4\rho\rho'}{[\rho+\rho']^2+[\zeta\mp\zeta_s(\rho')]^2}}
\end{eqnarray*}
in the expressions (\ref{Pois_Int2}) and (\ref{Pois_Int3}) 
respectively\footnote{Writing (\ref{Pois_Int3})
as a contour integral over a vector field and using 
Stoke's integral theorem as well as 
functional relations between elliptic integrals, one arrives at the
expression (\ref{Pois_Int2}).}.

Inserting the integral expression (\ref{Pois_Int3}) into the surface condition
(\ref{OFB}), one gets a non-linear integral equation for the unknown shape 
of the body's surface, which is uniquely soluble at least in the vicinity of 
some given solution if an appropriate triple of parameters is 
prescribed\footnote{As an example, the density $\mu$, the angular velocity 
$\Omega$ and the constant $\Phi_0$ of the fluid body can be taken. However, 
at the bifurcation points $\varepsilon_k$ it is necessary to use different 
parameters, in order to assure a unique solution, see below.}.
We write the shape of the body's surface parametrically in terms of a
Chebyshev expansion in the following manner:
\begin{eqnarray*}
   \rho_s(\tau)&=&\sqrt{\rho_1^2+(\rho_2^2-\rho_1^2)\tau}\;,\\
   \zeta_s[\rho_s(\tau)]&=&
      \sqrt{(1-\tau)[\zeta_1^2+\tau g(\tau)]}\;,\\
    g(\tau)&=&\sum_{k=1}^\infty g_k T_{k-1}(2\tau-1)-\frac{g_1}{2}\;,\\
    T_k(x)&=&\cos(k\arccos x).\end{eqnarray*}
The variable $\tau$ runs over the interval $[0,1]$, and the parameter $\zeta_1$
represents the polar radius for spheroidal 
figures and $\zeta_1=0$ for toroidal shapes. 

If we prescribe $\mu, \Omega$ and $\Phi_0$, restrict ourselves to a finite
number $(N-2)$ of coefficients $g_k$ 
and take the surface condition (\ref{OFB}) at the $N$ 
surface points $(\rho_s(\tau_k),\zeta_s[\rho_s(\tau_k)])$ with
$T_N(2\tau_k-1)=0$, the integral equation in question turns 
into a system of $N$ 
equations for the unknown coefficients $g_k$ and the likewise unknown
quantities $r_1$ and $\rho_2$ where
$r_1$ stands for $\zeta_1$ in the spheroidal and for $(-\rho_1)$ in the toroidal
case. 
Alternatively, we can 
prescribe the value $g(\tau=1)$ and search instead for $\Phi_0$. A choice
$g(1)\neq 0$ assures a unique solution at the
desired $\varepsilon_k$-sequence if we are in the vicinity of $\varepsilon_k$.
This is due to the fact, that for all
Maclaurin solutions $g(\tau)$ is identically zero. 

The solution of the remaining system of $N$ equations can be found by means of
a Newton-Raphson method. The 
Maclaurin ellipsoids at the bifurcation points serve as initial solutions, and
by choosing $g(1)<0$ we move towards the 
mass-shedding limit of the respective sequence whereas $g(1)>0$ yields the
path towards the ring or 
core-ring structures, see Section 4. The mass-shedding limit is reached when
$g(1)=-\zeta_1^2$, which leads 
to a cusp at the equatorial rim of the body. Although for this configuration 
derivatives of the above function $g$ are singular at $\tau=1$ 
(which results in a loss of accuracy, see below), we can choose 
$g(1)/\zeta_1^2$ as one of our parameters 
and can thus place ourselves precisely on this point.

Once upon the sequence and sufficiently far away from the Maclaurin sequence,
we 
may change the parameter prescription, e.g.~to the triples
($\mu,\Omega,\Phi_0$) or ($\mu,r_1/\rho_2,\Phi_0$). 
Note that rapid convergence of this method towards the actual shape of the
body as $N$ grows, requires, for a given 
set of coefficients $g_k$, a highly accurate evaluation of the integral
expression (\ref{Pois_Int3}).
This is done by dividing the interval $[0,1]$ into many subintervals, which are 
extremely
small in the vicinity of the respective zero $\tau_k$ and grow
exponentially as one moves away from $\tau_k$.
Moreover, for $m\approx 1$ the elliptic functions are written in logarithmic
expansions 
(see Gradstein \& Ryshik 1981, p. 295 f.). Additionally, special care is taken for surface points
close to the equatorial plane. 
In this manner we achieve an accuracy of up to 12 digits for intermediate
solutions that are sufficiently far away
from the mass-shedding limit as well as from any transition body where the
topology of the figure changes. The loss of accuracy as one approaches these
critical configurations comes about since derivatives of $g$ become singular 
at a certain point\footnote{This critical point is the equatorial rim for the
mass-shedding configuration, the `pinching' point for the configuration 
at which the two-body system forms and the origin $\rho=0=\zeta$ for the 
transition body of spheroidal to toroidal figures.} on the body's surface.
However, we are still able to obtain an accuracy of 5 digits for all physical
quantities given in Tables 2-5.

\begin{table}{{\bf Table 2:} Physical quantities for the
$\varepsilon_1$-configurations $(A)\ldots(L)$ displayed in Fig. 6.
The abbreviations `MSL' and `MLE' refer to
the mass-shedding limit and the Maclaurin ellipsoid respectively.}
\begin{center}
\begin{tabular}[b]{ccccl}
\hline\hline
$r_1/\rho_2$ & $\omega^2$ & $j^2$    & $T/|W|$ & \\
\hline
0.19224      & 0.092400   & 0.019847 & 0.34379 &  $(A)$: MSL \\
0.18500      & 0.089834   & 0.020815 & 0.35175 &  $(B)$ \\
0.17126      & 0.087262   & 0.021741 & 0.35890 &  $(C)$: MLE \\
0.16000      & 0.085614   & 0.022260 & 0.36249 &  $(D)$ \\
0.13000      & 0.082464   & 0.022927 & 0.36549 &  $(E)$ \\
0.10000      & 0.080648   & 0.022898 & 0.36260 &  $(F)$ \\
0.050000     & 0.079809   & 0.022215 & 0.35404 &  $(G)$ \\
0.000000     & 0.080181   & 0.021836 & 0.35047 &  $(H)$ \\
-0.20000     & 0.078338   & 0.020036 & 0.32825 &  $(I)$ \\
-0.40000     & 0.057176   & 0.021726 & 0.31082 &  $(J)$ \\
-0.60000     & 0.030207   & 0.031052 & 0.31542 &  $(K)$ \\
-0.80000     & 0.0089862 & 0.063778 & 0.33800 &  $(L)$
 \end{tabular}
\end{center}
\end{table}
\begin{table}{{\bf Table 3:} Physical quantities for the
$\varepsilon_2$-configurations $(A)\ldots(L)$ displayed in Fig. 7.
The abbreviations `MSL', `MLE' and `2-BS' refer to
the mass-shedding limit, the Maclaurin ellipsoid and the formation of the
two-body system respectively.}
% The final mass ratio of inner core and 
%outer ring amounts to $M_1/M_2=0.41051$.}
\begin{center}
\begin{tabular}[t]{ccccl}
\hline\hline
$r_1/\rho_2$ & $\omega^2$ & $j^2$    & $T/|W|$ & \\
\hline
0.11207      & 0.069411   & 0.027909 & 0.39710  &  $(A)$: MSL \\
0.11140      & 0.068335   & 0.028310 & 0.39919  &  $(B)$ \\
0.11160      & 0.066105   & 0.029152 & 0.40345  &  $(C)$: MLE \\
0.11300      & 0.064738   & 0.029663 & 0.40587  &  $(D)$\\
0.11500      & 0.063529   & 0.030101 & 0.40779  &  $(E)$\\
0.12000      & 0.061475   & 0.030787 & 0.41032  &  $(F)$\\
0.12500      & 0.059960   & 0.031218 & 0.41132  &  $(G)$\\
0.13000      & 0.058718   & 0.031496 & 0.41137  &  $(H)$\\
0.13500      & 0.057661   & 0.031659 & 0.41067  &  $(I)$\\
0.14500      & 0.055961   & 0.031701 & 0.40744  &  $(J)$\\
0.15500      & 0.054729   & 0.031417 & 0.40219  &  $(K)$\\
0.17558      & 0.054327   & 0.029720 & 0.38610  &  $(L)$: 2-BS\\
\end{tabular}
\end{center}
\end{table}
\begin{table}{{\bf Table 4:} Physical quantities for the
$\varepsilon_3$-configurations $(A)\ldots(L)$ displayed in Fig. 8.
The abbreviations `MSL', `MLE' and `2-BS' refer to
the mass-shedding limit, the Maclaurin ellipsoid and the formation of the
two-body system respectively. }
%The final mass ratio of inner and 
%outer ring amounts to $M_1/M_2=0.86563$.}
\begin{center}
\begin{tabular}{ccccl}
\hline\hline
$r_1/\rho_2$ & $\omega^2$ & $j^2$    & $T/|W|$ & \\
\hline
0.087646     &  0.054735  & 0.033738 & 0.42327 & $(A)$: MSL \\
0.086000     &  0.053836  & 0.034134 & 0.42477 & $(B)$\\
0.082750     &  0.052711  & 0.034638 & 0.42664 & $(C)$: MLE\\
0.075000     &  0.050871  & 0.035462 & 0.42947 & $(D)$\\
0.060000     &  0.048799  & 0.036307 & 0.43162 & $(E)$\\
0.040000     &  0.047554  & 0.036625 & 0.43116 & $(F)$\\
0.020000     &  0.047289  & 0.036544 & 0.42979 & $(G)$\\
0.000000     &  0.047391  & 0.036458 & 0.42932 & $(H)$\\
-0.050000    &  0.047666  & 0.036216 & 0.42800 & $(I)$\\
-0.10000     &  0.046297  & 0.036022 & 0.42281 & $(J)$\\
-0.12000     &  0.045352  & 0.035860 & 0.41895 & $(K)$\\
-0.13998     &  0.044952  & 0.035174 & 0.41220 & $(L)$: 2-BS\\
\end{tabular}
\end{center}
\end{table}

\begin{center}
\begin{figure}
\unitlength1cm
\hspace*{5mm}
\epsfig{file=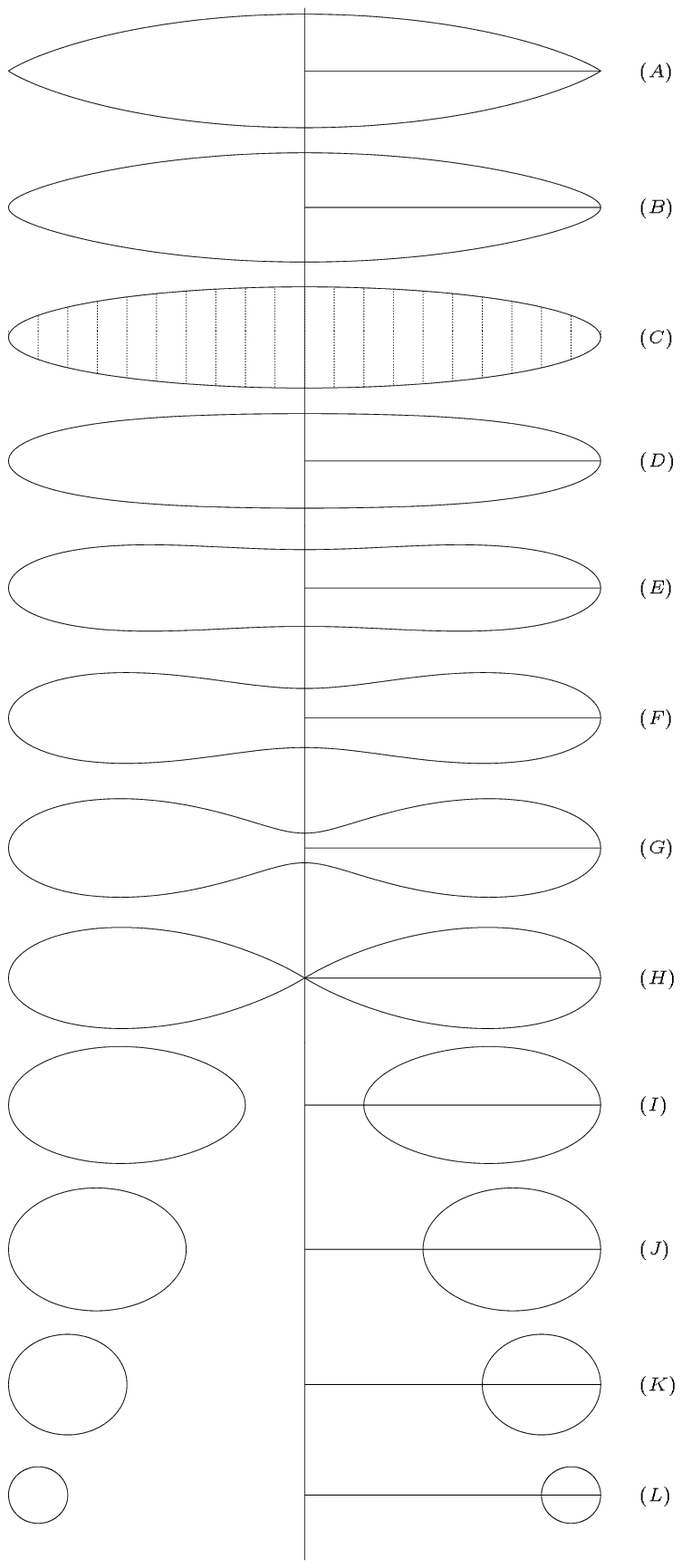,scale=1}
%\vspace*{5mm}
\caption{Meridional cross-sections of configurations belonging to the 
$\varepsilon_1$-sequence, see Table 2. The dimensionless quantity
$\zeta/\rho_2$ is plotted against $\rho/\rho_2$. The Maclaurin body is
denoted by hatching.}
\end{figure}
\end{center}

 \begin{center}
\begin{figure}
\unitlength1cm
\hspace*{5mm}
\epsfig{file=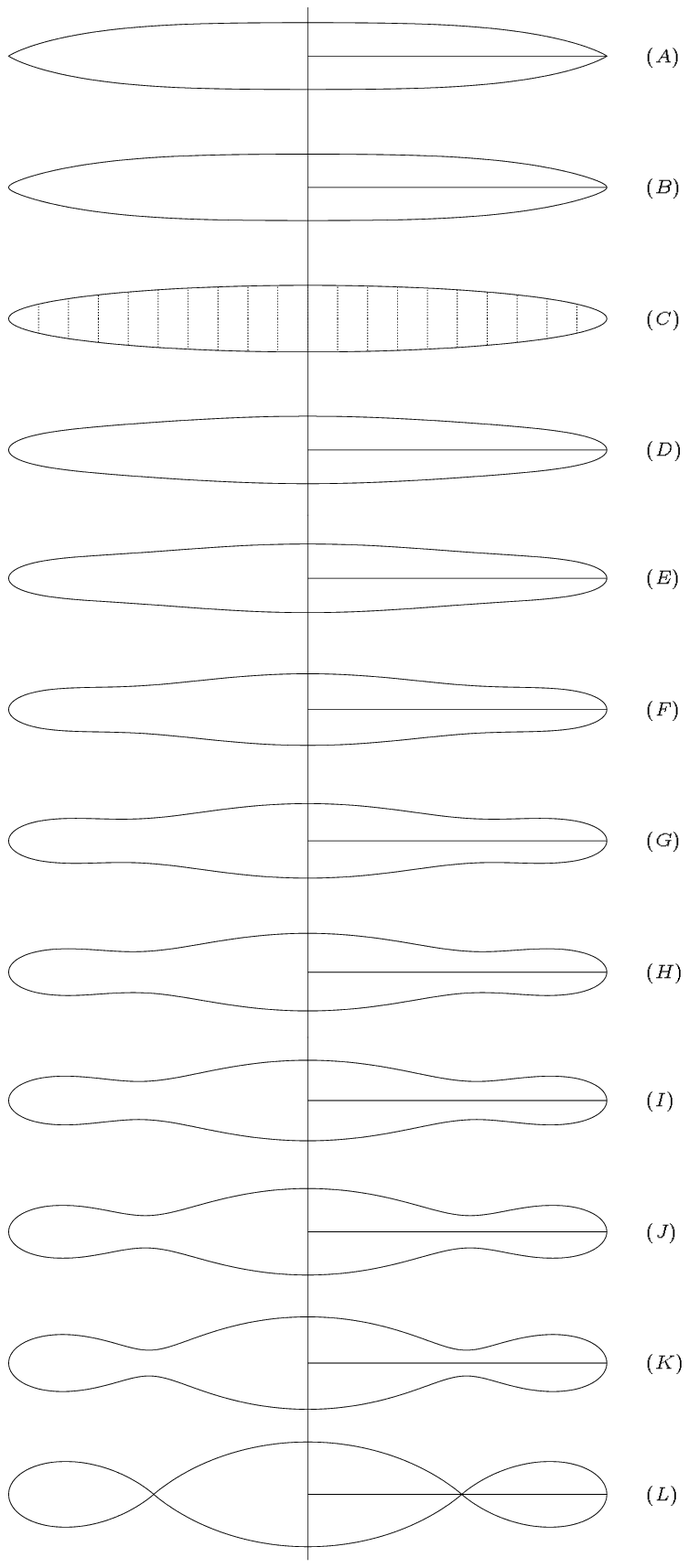,scale=1}
%\vspace*{5mm}
\caption{Meridional cross-sections of configurations belonging to the 
$\varepsilon_2$-sequence, see Table 3. The dimensionless quantity
$\zeta/\rho_2$ is plotted against $\rho/\rho_2$. The Maclaurin body is
denoted by hatching.}
\end{figure}
\end{center}

\begin{center}
\begin{figure}
\unitlength1cm
\hspace*{5mm}
\epsfig{file=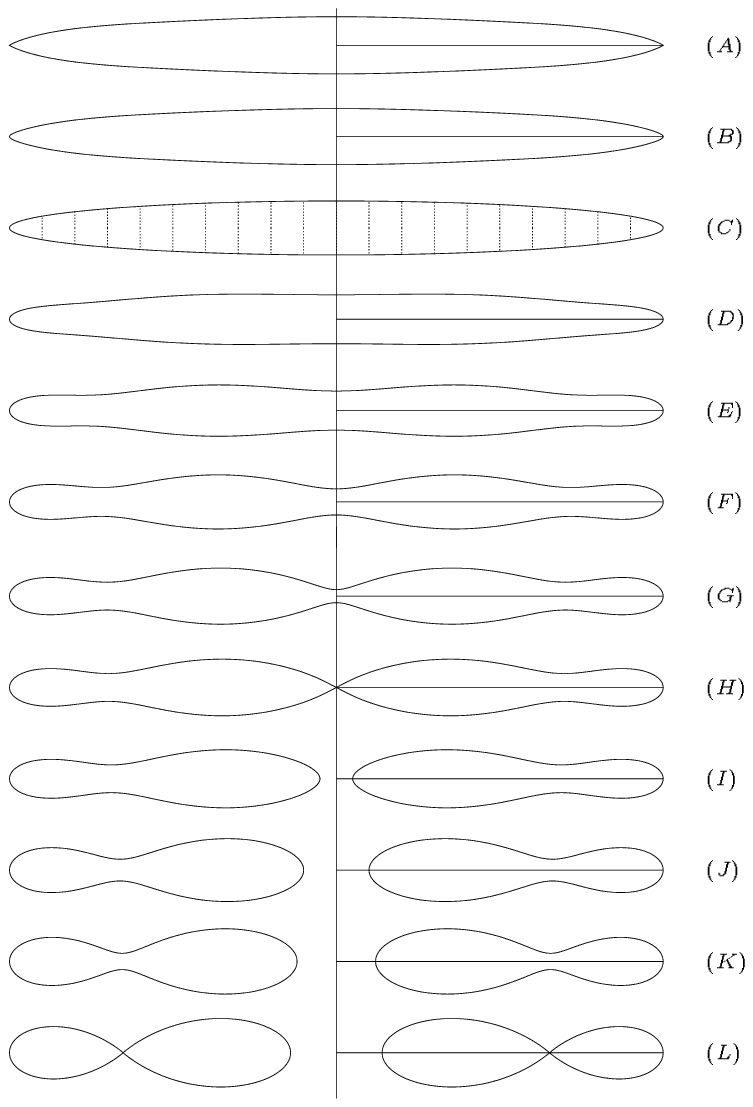,scale=1}
%\vspace*{5mm}
\caption{Meridional cross-sections of configurations belonging to the 
$\varepsilon_3$-sequence, see Table 4. The dimensionless quantity
$\zeta/\rho_2$ is plotted against $\rho/\rho_2$. The Maclaurin body is
denoted by hatching.}
\end{figure}
\end{center}

\begin{center}
\begin{figure}
\unitlength1cm
\hspace*{5mm}
\epsfig{file=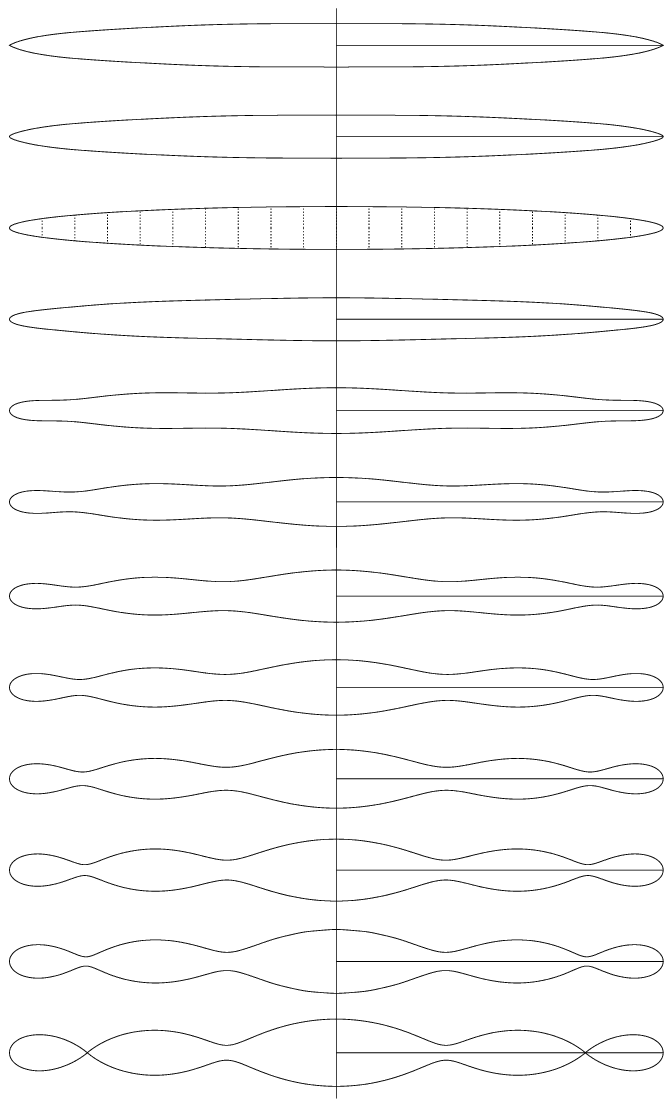,scale=1}
%\vspace*{5mm}
\caption{Meridional cross-sections of configurations belonging to the 
$\varepsilon_4$-sequence. The dimensionless quantity
$\zeta/\rho_2$ is plotted against $\rho/\rho_2$. The Maclaurin body is
denoted by hatching.}
\end{figure}
\end{center}

\begin{center}
\begin{figure}
\unitlength1cm
\hspace*{5mm}
\epsfig{file=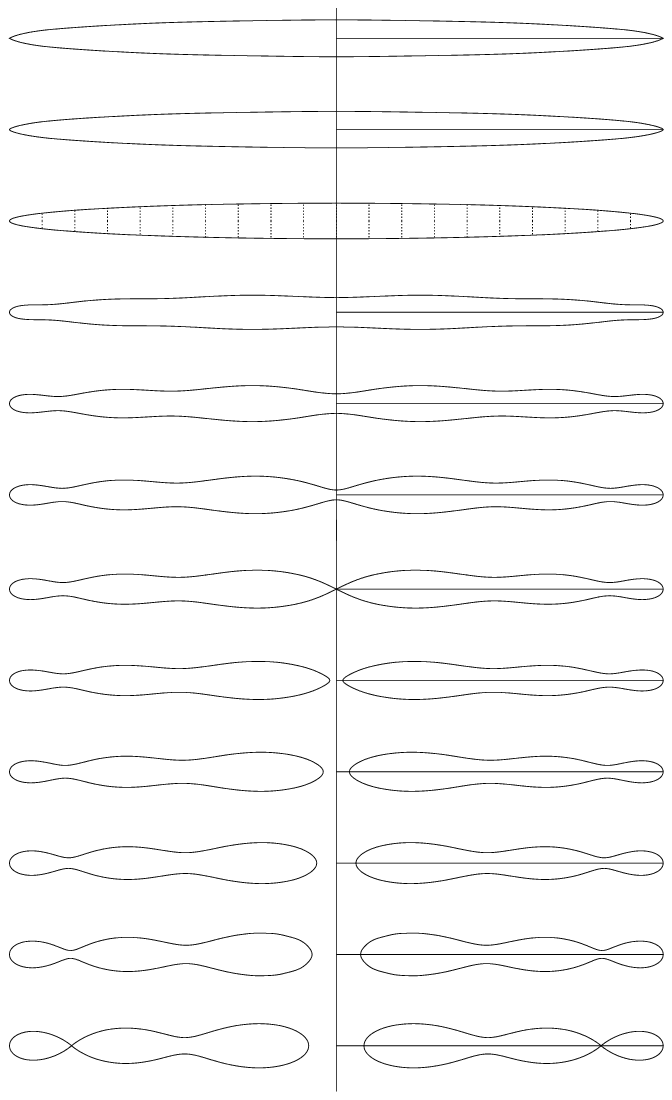,scale=1}
%\vspace*{5mm}
\caption{Meridional cross-sections of configurations belonging to the 
$\varepsilon_5$-sequence. The dimensionless quantity
$\zeta/\rho_2$ is plotted against $\rho/\rho_2$. The Maclaurin body is
denoted by hatching.}
\end{figure}
\end{center}
\begin{figure}
\unitlength1cm
\hspace*{-5mm}
\epsfig{file=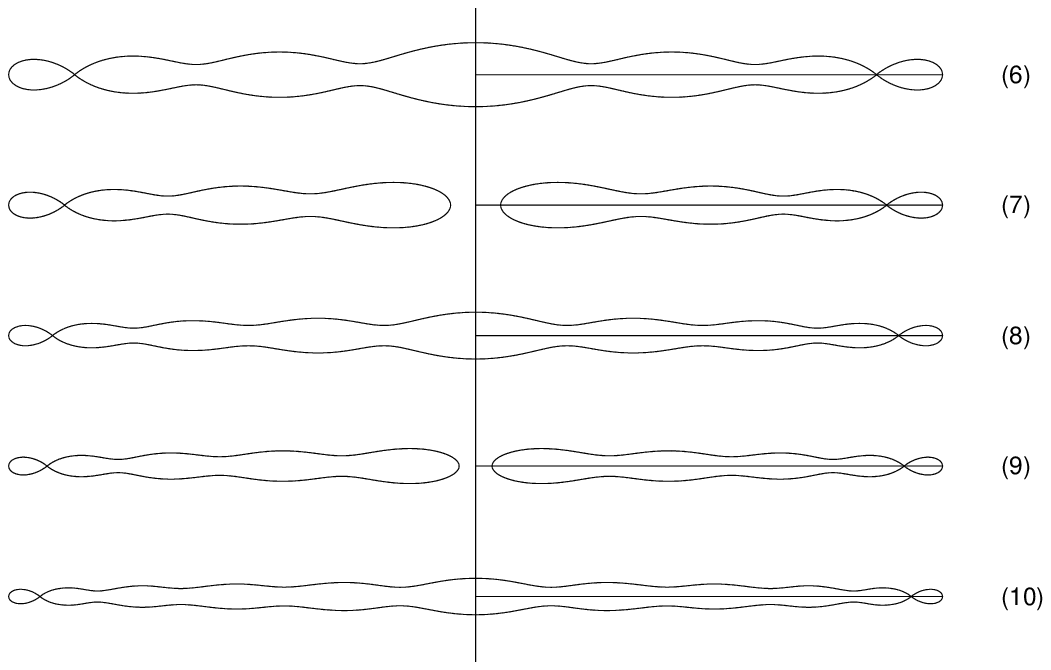,scale=0.7}
\vspace*{-10mm}
\caption{The formation of the two-body system at the end of the
$\varepsilon_6\ldots\varepsilon_{10}$-sequences.}
\end{figure}

%______________________________
\section{Results}

The evolutions of the five sequences bifurcating at the eccentricities
$\varepsilon_1,\ldots,\varepsilon_5$ 
are displayed in the Figs 1 and 2 where $\omega^2=\Omega^2/(4\pi G\mu)$ is
plotted
against the ratio $r_1/\rho_2$
and against the dimensionless angular momentum $j^2$ 
defined by
 \begin{equation}
 \label{j2}
 j^2=\frac{J^2\mu^{1/3}}{4\pi G M^{10/3}}
 \end{equation}
where $M$ and $J$ are the mass and the angular momentum of the body,
respectively. The corresponding (dotted) curves of the Maclaurin
ellipsoids are given as well as the 
analytic Dyson approximation of the $\varepsilon_1$-sequence (dashed lines).

In Fig. 1, the bifurcation points appear as cross-over points of the Maclaurin
and respective $\varepsilon_k$-sequence. One end of a particular sequence
is marked by a mass-shedding limit, always possessing a higher angular
velocity $\omega^2$ compared to the other end. The details of the typical 
evolution of a particular sequence can be seen in Fig. 3.
A characteristic feature of all sequences is the occurence of a local maximum
in $\omega^2$ close to the mass-shedding limit, see Figs 3(b), 3(c) and 4(b).

All $\varepsilon_{2l-1}$-sequences 
show the transition of a spheroidal to a 
toroidal figure, and $\omega^2$ possesses a local 
minimum and maximum before and after this point respectively, see also 
Figs 3(a) and 4(a). 
Another typical property is the vanishing gradient of 
$\omega^2$ as a function of $r_1/\rho_2$ as well as a discontinuous change of 
the sign of its curvature at $r_1/\rho_2=0$. Apart from the
$\varepsilon_1$-sequence, the paths end with the
formation of a two-body system, namely a core and a ring for $k=2l$ and
two rings if $k=2l+1$. Shortly before coming to this end, a minimum in 
$\omega^2$ occurs, which becomes less and less pronounced as one moves 
to higher values of $k$.

As depicted by Hachisu \& Eriguchi (1983), the paths of the 
$\varepsilon_k$-sequences glance off the Maclaurin 
sequence\footnote{At the bifurcation 
points the slopes are identical.},
if one plots $\omega^2$ against $j^2$, see Fig. 2.
Representative details corresponding to the $\varepsilon_1$-curve 
can be seen in Fig. 4. 
Note the extraordinary accuracy needed in order to resolve
the global maximum in $\omega^2$ close to the mass-shedding limit displayed
in Fig. 4(b). 

In Figs 1 and 2, moreover the analytic curves of the Dyson approximation
can be found. With the assumption of circular cross-sections of the toroids as
$\rho_1/\rho_2\to 1$, one may expand (\ref{Pois_Int2}) in terms of
$(1-\rho_1/\rho_2)$ and find by means of (\ref{OFB}) an expansion of the
toroids' surfaces. Dyson went to the order $o[(1-\rho_1/\rho_2)^4]$,
resulting in excellent agreement with the numerical results\footnote{In 
our computations, we compared configurations 
with the same $\rho_1,\rho_2$ and $\mu$.}
for large $\rho_1/\rho_2$.

Fig. 5(a) is a replot of fig. 1 from Bardeen (1971), with the Dyson
approximation being replaced by the actual evolution of the
$\varepsilon_1$-sequence. Here, the dimensionless total energy $\bar{E}$ is
given by
\begin{equation}
\label{barE}
\bar{E}=\frac{T+W}{G\mu^{1/3}M^{5/3}}=2\pi j\omega\,(1+\frac{W}{T})
\end{equation}
with the kinetic and gravitational energies 
 \[ T=\frac{1}{2}J\Omega\quad\mbox{and}\quad 
 W=\frac{\mu}{2}\int\limits_V \Phi\,d^3{\bf
 r}=\frac{3}{5}M\Phi_0+T.\]
The latter  formula follows from the virial theorem 
\[2T+W+3\int\limits_V p\,d^3{\bf r}=0\;,\]
which is valid for all stationary rotating fluid bodies, and from the
hydrodynamical
Euler
equation giving for a uniformly rotating
homogeneous configuration the pressure $p$ by
\[p=\mu(\Phi_0-\Phi+\frac{1}{2}\Omega^2\rho^2)\;.\] 

Fig. 5(a) led Bardeen to the conjecture that the Dyson-ring sequence
branches off from the Maclaurin sequence exactly at the
$\varepsilon_1$-bifurcation point, marked in Fig. 5 by $\bullet$. However, a 
magnified detail of the plot in question shows a completely different picture,
as can be seen in Figs 5(b) and 5(c). Coming from the toroidal end, the
$\varepsilon_1$-sequence intersects the Maclaurin sequence close to but not
exactly at the bifurcation point. As can be seen in Fig.
5(c), the $\varepsilon_1$-sequence intersects the Maclaurin sequence another
time before it moves on to the bifurcation point and beyond to the 
mass-shedding limit\footnote{The relation $d\bar{E}=4\pi\omega\, dj$, which
is valid along any equilibrium sequence of uniformly rotating 
homogeneous
fluid bodies (cf. Hartle \& Sharp 1967, Bardeen 1971), explains two features of 
Fig. 5: the two cusps and the
fact that the slope of the $\varepsilon_1$-sequence is identical to that of
the Maclaurin sequence at the bifurcation point. Moreover, together with the
coinciding slopes in the $\omega^2$-$j^2$-plane (Fig. 2), one can conclude that
the second derivatives in the $\bar{E}$-$j^2$-plane (Fig. 5) are identical at
the bifurcation point as well.}.
It is rather surprising that from this entangled picture the
correct connections of the Dyson-ring and the Maclaurin sequence could be
drawn.

Table 2 contains numerical data, accurate to five digits, 
for physical quantities belonging to the $\varepsilon_1$-sequence. The
corresponding meridional cross-sections of these bodies are displayed in Fig. 6.
The intermediate Maclaurin body 
$(C)$ is denoted by hatching. Note that a lens-shaped
configuration $(B)$ appears that looks similar to the relativistic 
homogeneous fluid body displayed in fig. 1 (a) of Ansorg, Kleinw\"{a}chter \&
Meinel (2002).
Additionally, the values of physical quantities like $j^2$ and $\omega^2$ 
are similar for the 
two bodies\footnote{In Ansorg et al. (2002), $j^2=R/(4\pi)$ and
$\omega^2=\bar{\Omega}^2/(4\pi)$.}. 
Such a neighbourhood can also be found for the configurations $(F)$ and 
fig. 1 (b) of Ansorg et al. (2002).

Likewise, Tables 3 and 4 list data for the fluid bodies shown in Figs 7 and 
8. Again, there is a configuration $(B)$ of the $\varepsilon_2$-sequence (see
Fig. 7) which is in the vicinity of a relativistic body, (c) in fig. 1 of 
Ansorg et al. (2002).
Note the hint of a minimum at the north pole of the relativistic
figure. Also for the configuration $(B)$, this property is suggested by the 
linear surface displacement (\ref{surf_dis}).
But, as can be seen in Figs 6-10, the Newtonian
configurations in between the Maclaurin body and the mass-shedding limit 
neither generate a central bulge (which one would expect for the
$\varepsilon_{2l-1}$-sequences) nor a central minimum region (which would follow
for the $\varepsilon_{2l}$-sequences). However, in the general relativistic
vicinity of these solutions, at least the latter property seems to show up. 

Table 5 lists the ratios of the masses of the inner to the 
outer body at the two-body formation point for the
$\varepsilon_2\ldots\varepsilon_5$-sequences.

Figs 9 and 10 display configurations of the $\varepsilon_4$ and
$\varepsilon_5$-sequences, and Fig. 11 is devoted to the formation of 
the final two-body-system for the 
$\varepsilon_6\ldots\varepsilon_{10}$-sequences. As mentioned in the
Introduction, for $\varepsilon_{2l}$ the resulting configuration consists of
a corrugated central core-region and an outer ring which just
pinches off.
Likewise, for $\varepsilon_{2l+1}$ a two-ring system is formed. 

An interesting open question concerns the behaviour of the 
$\varepsilon_k$-sequences as $k$ tends to infinity, corresponding to the disk 
limit ($\varepsilon\to 1$) of the Maclaurin ellipsoids.
\begin{table}{{\bf Table 5:} The ratios $M_1/M_2$ of the masses of the inner to
the outer body at the two-body formation point for the
$\varepsilon_2\ldots\varepsilon_5$-sequences.}
\begin{center}
\begin{tabular}{ccccc}
\hline\hline
& (2) & (3) 
 & (4) & (5)\\
\hline
$M_1/M_2$ & 0.41051  & 0.86563 & 1.7235 & 2.4987 
\end{tabular}
\end{center}
\end{table}

\section*{Acknowledgements}
The authors would like to thank David Petroff for many valuable discussions. 
This work was supported by the German {\it Deutsche Forschungsgemeinschaft}
(DFG-project ME 1820/1).

\end{document}